\documentclass[iop,apj]{emulateapj}

\usepackage{amsmath}
\usepackage{mathrsfs}
\usepackage{color}

\usepackage{times}
\usepackage{color}

\def\bfbeta{\beta \kern-.680em \beta}

\def\bfomega{\omega \kern-.60em \omega}

\def\bfpsi{\psi \kern-.645em \psi}

\def\bfrho{\rho \kern-.55em \rho}

\def\bftau{\tau \kern-.524em \tau}

\def\bfzeta{\zeta \kern-.50em \zeta}

\def\bfkappa{\kappa \kern-.50em \kappa}

\def\bfdel{\nabla \kern-.770em \nabla}

\shorttitle{Ponderomotive Acceleration in Coronal Loops}
\shortauthors{Dahlburg, Laming, Taylor and Obenschain}

\begin{document}

\title{PONDEROMOTIVE ACCELERATION IN CORONAL LOOPS}

\author{R. B. Dahlburg$^{1}$, J. M. Laming$^2$, B. D. Taylor$^3$ and K. Obenschain$^1$}
\affil{\vspace{.2em}
$^1$LCP\&FD, Naval Research Laboratory, Washington, DC 20375, USA \\
$^2$Space Science Division, Naval Research Laboratory, Washington, DC 20375, USA
$^3$AFRL Eglin AFB, Pensacola, FL 32542, USA}

\begin{abstract}
Ponderomotive acceleration has been asserted to be a cause of the First Ionization Potential (FIP) effect,
the by now well known enhancement in abundance by a factor of 3-4 over photospheric values of elements
in the solar corona with FIP less than about 10 eV.
It is shown here by means of numerical simulations that ponderomotive acceleration occurs in
solar coronal loops, with the appropriate magnitude and direction, as a ``byproduct'' of
coronal heating.
The numerical simulations are performed with the HYPERION code, which solves the fully compressible three-dimensional magnetohydrodynamic equations including nonlinear thermal conduction and optically thin radiation. Numerical simulations of a coronal loops with an axial magnetic field from  0.005 Teslas to 0.02 Teslas and lengths from 25000 km to 75000 km are presented.  In the simulations the footpoints of the axial loop magnetic field are convected by random, large-scale motions.  There is a continuous formation and dissipation of field-aligned current sheets which act to heat the loop.  As a consequence of  coronal magnetic reconnection, small scale, high speed jets form.  The familiar vortex quadrupoles form at reconnection sites.  Between the magnetic footpoints and the corona the reconnection flow merges with the boundary flow.  It is in this region that the ponderomotive acceleration occurs.
Mirroring the character of the coronal reconnection, the ponderomotive acceleration is also found to be intermittent.

\end{abstract}


\section{Introduction}
Magnetohydrodynamic waves and their interactions with plasma and magnetic fields have many
manifestations in solar and stellar atmospheres. Besides the well-known coronal heating,
X-ray emission and other forms of activity, the difference between the solar or stellar
coronal composition and that of the underlying photosphere possibly represents one of the
most direct observables of processes generating magnetic field and waves. In solar conditions in
the chromosphere, the ponderomotive force due to Alfv\'enic waves usually draws chromospheric ions,
e.g. Fe, Si, Mg up into the corona, leaving the neutrals behind since the waves are fundamentally
oscillations of the magnetic field that interact with ions.  This phenomenon is called the
First Ionization Potential (FIP) effect. The FIP effect produces a coronal enhancement
in abundance by a factor of 3-4 over photospheric values of elements with FIP less than
about 10 eV  \citep[e.g.][]{feldman00}. The FIP effect is also observed in many solar-like late type stars. At higher
activity levels and/or later spectral types, a so called ``Inverse FIP'' effect is observed,
where the low FIP elements are depleted in the corona \citep[e.g.][]{wood10,doschek15}.

Previous work has shown that the FIP effect and many of its variations in different regions of the solar corona and
wind can all be explained with the ponderomotive force \citep{laming04,laming09,laming12,laming15}. In appropriate
conditions, the Inverse FIP effect can be modeled as well \citep{wood13,laming15}, and in fact
the ponderomotive force is the only model of the FIP effect that can also give rise to Inverse FIP.
An upwards ponderomotive acceleration of order $10^4$ m s$^{-2}$ is typically required at the steep chromospheric
density gradient to provide the fractionation. The waves responsible for this ponderomotive acceleration may either
be generated in the corona,
impinging from the corona to the upper chromosphere before being reflected back into the corona, or derive ultimately
from the mode conversion of photospheric p-modes. In this last case, waves impinging on the
chromosphere from below may either be transmitted upwards, or reflected back down into the solar envelope,
giving rise to the FIP or Inverse FIP effects respectively. The FIP fractionations produced at the
top of the chromosphere by reflecting coronal waves, or lower down by photospheric waves, are
broadly similar, but with subtle differences, for example in the depletion of He and Ne with
respect to O, that strongly favor the
coronal waves as the agent of fractionation producing the FIP effect seen in the majority of the solar
corona and wind. Evidence for such coronal waves can be found in the non-thermal mass motions
inferred from spectral line widths \citep[e.g.][]{baker13},
but they and the processes that generate them have yet to be convincingly
directly observed.

Accordingly in this paper we study numerical simulations of coronal heating to evaluate the
ponderomotive acceleration developing at loop footpoints, with a view to developing the FIP effect as a
diagnostic for MHD waves in the solar atmosphere. We concentrate on the {more prevalent
FIP effect produced by coronal waves.
The numerical simulations employ the 3D compressible MHD HYPERION code \citep{2010AIPC.1216...40D,
2012A&A...544L..20D, 2015ApJ...submitted} to determine whether or not  ponderomotive acceleration develops in a standard quiescent corona scenario. HYPERION is a parallelized Fourier collocation finite difference code with Runge-Kutta time discretization that solves the compressible MHD equations with parallel thermal conduction and radiation included.
In this context it is difficult to perform an analysis based on waves, which must be filtered out of a chaotic background.
Rather, in this paper we develop an interpretation based on magnetohydrodynamic turbulence theory.
We find that ponderomotive acceleration occurs naturally in our 3D compressible MHD simulations of
coronal heating using the HYPERION code. The coronal magnetic field is gradually stressed by
footpoint motions, and the built up energy is released in explosive magnetic reconnection events.
Hence the flow field at the magnetic footpoints is determined by convection, while the flow field in the
corona is determined by magnetic reconnection. Ponderomotive force occurs in the region where
the photospheric flow field transitions into the coronal flow field, which is also where the background
density gradient is very steep as the chromosphere transitions into the corona.

In Section 2 we describe the governing equations, the boundary and initial conditions, and the numerical method. In Section 3 we detail our numerical results.
We present details of the evolution and origins of the ponderomotive acceleration for several coronal loops. Loops with different lengths and magnetic field strengths are simulated and analyzed.  Section 4 contains a wave-based discussion of the results.  Section 5 presents our conclusions.

\section{Formulation of the problem} \label{sec:fp}
We use the HYPERION code to model the solar corona.
The HYPERION code is a computational representation of the Parker model that has been used to investigate coronal heating
\citep{1972ApJ...174..499P,1988ApJ...330..474P, 1994ISAA....1.....P}.
In this model the corona is represented as a square cylinder (a figure which represents the computational box is discussed in Section 3.3) .
The $x$ and $y$ directions are periodic.
The boundaries in the $z$ direction represent the upper chromosphere.
A DC magnetic field is applied along the $z$ direction.
The footpoints of this magnetic field are convected by applied flows at each $z$ boundary.
We model the solar corona as a compressible, viscoresistive magnetofluid with nonlinear parallel thermal conduction and optically thin radiation losses.
The interior of the computational box represents the corona.
We model the corona as a compressible, viscoresistive magnetofluid with nonlinear thermal conduction and optically thin radiation losses.
The equations which govern this system are given in the Appendix.

\subsection{Initial and boundary conditions} \label{sec:inibc}

We solve the governing equations in a Cartesian domain of size
$L_x \times L_y \times L_z = 1 \times 1 \times L_z$,
where $L_z$ is the loop aspect ratio determined by the loop length
and the characteristic length ($0\le x,y \le1,\ -L_z/2 \le z \le L_z/2$).
The system has periodic boundary conditions in $x$ and $y$,  line-tied boundary
conditions at both $z$ boundaries,
and it is threaded by a strong guide magnetic field $B_0 = 1$ in the $z$-direction.

To convect the magnetic footpoints
at the boundary $z=L_z/2$ boundary we evolve the stream function:
\begin{equation} \label{eq:fc1}
\psi_t (x, y, t) = \xi_1 \sin^2 \left(\frac{\pi t}{2 t^*}\right) + \xi_2  \sin^2 \left(\frac{\pi t}{2 t^*} + \frac{\pi}{2}\right),
\end{equation}
and at the boundary $z=-L_z/2$ boundary we evolve the stream function:
\begin{equation} \label{eq:fc2}
\psi_b  (x, y, t) = \xi_3 \sin^2 \left(\frac{\pi t}{2 t^*}+ \frac{\pi}{4}\right) + \xi_4  \sin^2 \left(\frac{\pi t}{2 t^*} + \frac{3\pi}{4}\right),
\end{equation}
where
\begin{equation} \label{eq:fc3}
\xi_i (x,y) =  \sum_{n, m} \frac{a_{nm}^i\ \sin \left[2\pi (n x + m y + \zeta_{nm}^i) \right]}
                 {2\pi \sqrt{n^2 + m^2} },
\end{equation}
in which all wave-numbers with $3\le \sqrt{n^2 + m^2} \le4$ are excited, so that
the typical length-scale of the eddies is $\sim 1/4$
(where $\mathbf{v} = \nabla  \psi \times \mathbf{\hat e}_z$)
\citep{2015ApJ...submitted}
The characteristic length in our simulations is chosen to be 4000 km.   The wavenumber restriction prioritizes wavelengths of approximately 1000 km, which is approximately the size of a typical photospheric granule.).
$a_{nm}^i$ and $\zeta_{nm}^i$ are random numbers chosen such that
$0\le a_{nm}^i, \zeta_{nm}^i \le1$.
Every $t^*$, the coefficients $a_{nm}^i$ and  $ \zeta_{nm}^i$ are randomly
changed alternatively for eddies 1 through 4.
The magnetic field is expressed as
$\mathbf{B} = B_0 \mathbf{\hat e}_z + \mathbf{b}$
with $\mathbf{b} (x,y,z,t)=  \nabla \times \mathbf{A}$, where $\mathbf{A}$ is the vector
potential associated with the fluctuating magnetic field.
At the both $z$ boundaries $B_z$, $n$ and $T$
are kept constant at their initial values $B_0$, $n_0$ and $T_0$,
while the magnetic vector potential is convected by the resulting flows.
The initial number density and temperature profiles are determined by the elliptical gravity model
\citep{2015ApJ...submitted}.

In what follows we have assumed the normalizing quantities to be:
$n_*=10^{17}\,$ m$^{-3}$,
$T_*=10^{4}\,$  K, and $L_* = 4 \times 10^{6}\,$  m.
We set the Coulomb logarithm ($\ln \Lambda$) equal to 10.
A basic loop length of $L_{z*}= 6.25~L_*$= 25,000 km is used.  Simulations with $2L$  and $3L$ were also performed.
In this paper the cases with lengths $L,~2L,$ and $3L$ are referred to as case A, case B, and case C respectively (see Table 1).
A basic loop magnetic field strength of $B_*=0.01$ Teslas is used. Simulations with half and twice this magnetic field strength
are also performed (see Table 1).
The normalized time scale of the forcing, $t^*$, is set to a 5 minute convection time scale. Thus the normalized
driving velocity $V_*$ is $10^3$~m~s$^{-1}$.

\bigskip\bigskip\bigskip\bigskip\bigskip\bigskip

\subsection{Ponderomotive acceleration}\label{sec:pma}

Ponderomotive acceleration has been shown to be a possible cause of the FIP effect
\citep{laming04,laming09,laming12,laming15}.
In dimensionless form the dimensionless ponderomotive acceleration is given by:
\begin{equation}
a=0.5 {\partial\over\partial s}\Bigl[ {(\delta{\bf E})^2\over {\bf B}^2} \Bigr]
\label{paeq}
\end{equation}
Here $\delta{\bf E}$, the fluctuating electric field, can be evaluated as:
\begin{equation}
\delta{\bf E} = - {\bf v}\times{\bf B} + {1\over S} \nabla\times {\bf B}
\label{eeq}
\end{equation}
where ${\partial\over\partial s}$ denotes a spatial derivative taken along the magnetic field line.
For the ponderomotive acceleration to be significant in producing the FIP effect, it must
act in an appropriate spatial direction to counteract the effects of solar surface gravity ($g$).
In the context of our model the ponderomotive acceleration should be directed away from the $z$-boundaries (upper chromosphere) and toward the
$z$-midplane (loop apex) of the computational box.
For the typical loop, the axial magnetic field should be approximately constant.
Equation \ref{paeq} then implies that, to obtain the FIP effect, the $s$ derivative of the square of the fluctuating electric field should be positive close to the $z=-L_z / 2$ boundary or negative near the $z=L_z / 2 $ boundary.

Note as well that there are two terms which contribute to the fluctuating electric field (equation \ref{eeq}):
an ideal component $(- {\bf v}\times{\bf B})$ and a resistive component $({1\over S} \nabla\times {\bf B})$.
In what follows we will examine how these two components vary along a coronal loop.
A further point to note is that it is the perpendicular velocity and fluctuating magnetic fields that are significant for the formation of the fluctuating electric field.  For the ideal part of the electric field, the largest components are formed by the cross product of the perpendicular velocity components and the large axial magnetic field.  For the resistive part of the electric field, the perpendicular fluctuating magnetic field is responsible for the formation of the large current sheets directed along the axial magnetic field.
In the results section of this paper we will examine these fields to see if they exhibit the needed behavior.

\section{Results} \label{sec:res}

\subsection{Some definitions} \label{sec:sensor}
In this section we will outline a statistical analysis of the turbulent behavior of the loop, with an emphasis on quantities that help to
delineate features of the ponderomotive acceleration.
Parameters for the simulations are found in Table 1.
To elucidate the statistical approach it is helpful to imagine that there is a sensor at each grid point of our simulated coronal loop.
This sensor could measure, for example, the number density ($n$).
To perform the analysis we then average $n$ over the horizontal planes at some specified temporal frequency.
As an example the data could be horizontally averaged once every 100 time steps.
All of these horizontal averages are then time averaged over some specified temporal interval.
The result of all of this averaging is a curve in $z$ that allows us to interpret aspects of the behavior of the desired function, in this example $n$.
For example, we can determine the average number density as a function of $z$.
To gain fuller insight into the relation of ponderomotive acceleration to the behavior of the turbulent functions, it will also be necessary to compute  higher order statistical functions such as the skewness and kurtosis.

As an example of how we perform the statistical analysis,
let $<f>$ denote the time average of a function $f$ which has been spatially averaged over the perpendicular directions, {\it i.e.,}
\begin{equation}
<f> = {1\over (m_f - m_i)}\sum_{m=m_i}^{m_f}\Bigl[{1\over{n_x}{n_y}}\sum_{j=1}^{n_y}\sum_{i=1}^{n_x} f_{ijm}\Bigr]
\end{equation}
where $f_{ijm}= f(x_i,~y_j,~z,~t_m)$.
Here $m$ indexes the time interval for averaging, with $m_f > m_i$.
This will give the average for mean quantities, such as the mean temperature or number density.
For example, the averaged mean mass density will be given by:
\begin{equation}
<n (z)>= {1\over (m_f - m_i)}\sum_{m=m_i}^{m_f}\Bigl[{1\over{n_x}{n_y}}\sum_{j=1}^{n_y}\sum_{i=1}^{n_x} n_{ijm}\Bigr].
\end{equation}
Typically the temporal averaging is performed for at least 1800 seconds of physical time.

\subsection{Evidence of ponderomotive acceleration}
The footpoints of the magnetic field are subjected to convection at the $z$ boundaries with the initial number density and temperature profiles stratified as described in subsection \ref{sec:inibc}.
The kinetic energy of the footpoint driving motions is transformed into magnetic energy as the loop magnetic field lines are stretched and twisted.
Most of this magnetic energy is then converted into thermal energy and kinetic energy by means of magnetic reconnection.
The thermal energy is efficiently conducted to the $z$ boundary region where radiation is effective.
This energization takes some time, so that to avoid an initial catastrophic cool-down the thermal conduction and optically thin radiation are ramped up
from zero to their final values over about five minutes of dimensioned time.
It takes approximately 100 Alfv\'en times (ten minutes of dimensioned time) to achieve a steady state, at which time we begin to sample the fields for the statistical analysis.

Before proceeding to the analysis of ponderomotive acceleration, it is instructive to examine the thermodynamic features of the loop.
Figure \ref{ntmean} shows the averaged number density and temperature for run B.
\begin{figure}
  \includegraphics[width=1.\columnwidth, bb = 150 40 635 500]{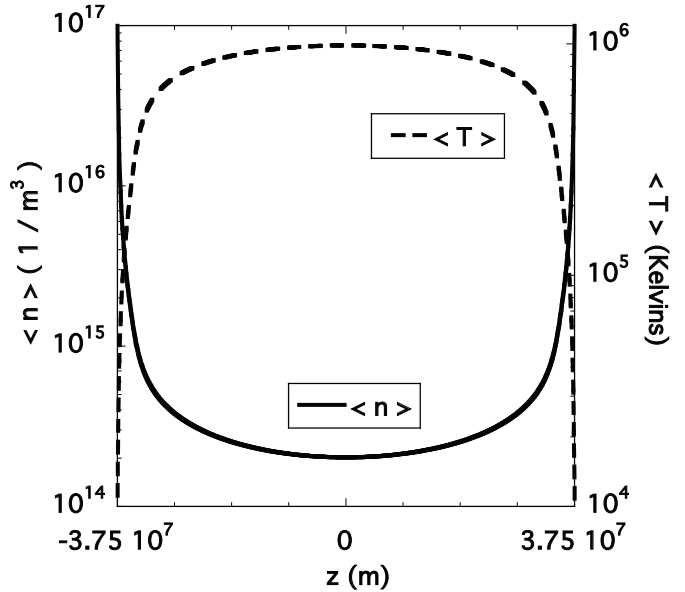}
  \caption{Average number density and temperature {\it vs} $z$ for case C.}
\label{ntmean}
\end{figure}
Note that the averaged temperature at the loop apex is lower than the prescribed initial value of $10^6$ Kelvins.
Hence on average the loop is cooling down rather than heating.
However, an analysis of fluctuations about this state shows that unsteady heating occurs.
Figure \ref{ntrms} shows the root mean square variation of the number density and temperature for run B.
Note that at the loop apex there are RMS temperature fluctuations of about $2\times 10^5$ Kelvins.
\begin{figure}
  \includegraphics[width=1.\columnwidth, bb = 150 40 635 500]{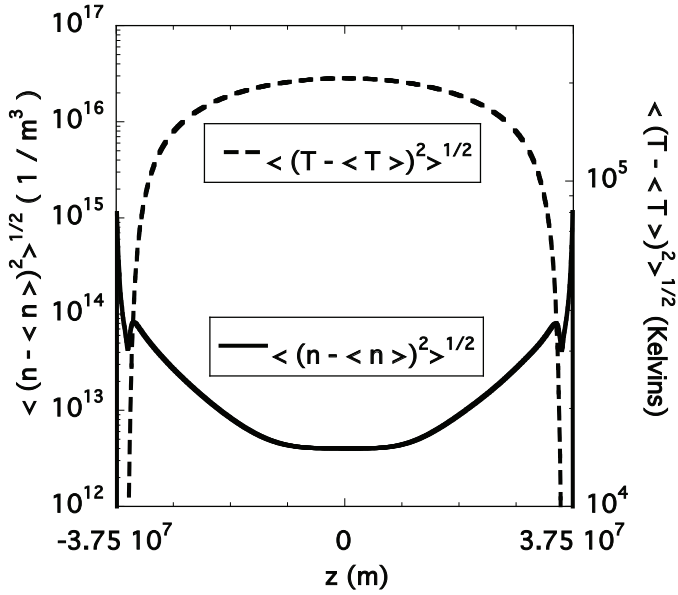}
  \caption{Root mean square variation of number density and temperature {\it vs} $z$ for case C.}
\label{ntrms}
\end{figure}
The dynamics of the system are responsible for this heating.

This statistically steady state, which we have extensively analyzed elsewhere \citep{2010AIPC.1216...40D,
2012A&A...544L..20D, 2015ApJ...submitted}, is characterized by intermittent MHD turbulence. There is a chaotic release of energy due to magnetic reconnection in the central part of the channel. While most interest has centered on the transformation of magnetic energy into heat, there is also a significant transformation of magnetic energy into kinetic energy. The morphological consequence of this transformation is the development of reconnection jets in the flow. These are high speed, highly localized structures.
In contrast the driving flows at the wall are low speed, large spatial scale structures (see subsection 2.2).
Hence there is a mismatch between the wall flow and the internal flow.
This mismatch can provide the gradient required to produce ponderomotive acceleration (see equations \ref{paeq} and \ref{eeq}).

The ponderomotive acceleration depends on the parallel spatial derivative of the square of the electric field
divided by the magnetic fleld. As noted before, the electric field has an ideal component $(- {\bf v}\times{\bf B})$ and
a resistive component $({1\over S} \nabla\times {\bf B})$.
As explained in section \ref{sec:pma}, the perpendicular components of the velocity and fluctuating magnetic field are the most significant.
For the ponderomotive force to be active, there must be a variation of these components from the $z$ boundary to the center
of the system.
One way to examine this is to look at the perpendicular velocity and magnetic field average two-point correlations with respect to the $z$ coordinate, {\i.e.,}
approximately along the direction of the loop magnetic field.
For a function $f$ we define the average two-point correlation with function value at $z = 0$ in the following way:
\begin{equation}
C(f)={<f(x,y,z)> <f(x,y,z+\Delta z>\over<f^2(x,y,z)>}
\end{equation}
where $\Delta z$ is the grid spacing in the $z$ direction.
The two-point correlations for the perpendicular dynamic field components are shown in Figure \ref{ubcorr}, where the reference fields are taken at the left boundary.
Note that both fields decorrelate toward the center of the channel, but in general the perpendicular velocity field fluctuations lose contact with the wall values more rapidly than the perpendicular magnetic field fluctuations.
The high correlation shown in the perpendicular magnetic field components is reflective of the formation of highly elongated parallel current sheets.
The decorrelation in the perpendicular velocity field occurs because the eddies formed at the wall by forcing are transformed into reconnection jets in the channel center.
This implies that the important physics for the ponderomotive acceleration involves the velocity field.

\begin{figure}
  \includegraphics[width=1.\columnwidth, bb = 150 40 635 500]{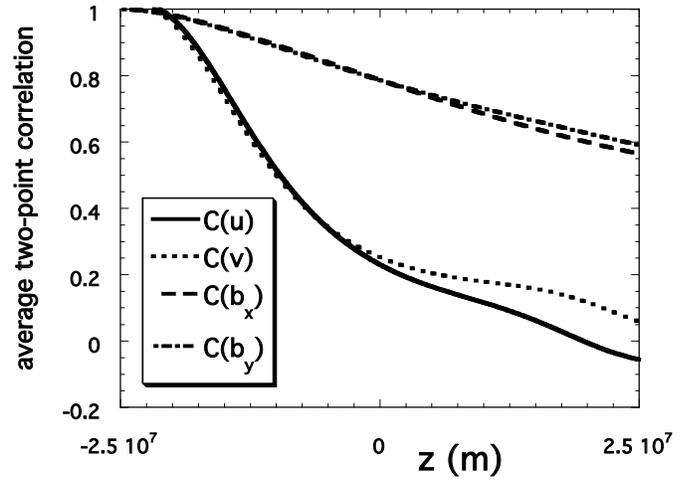}
  \caption{Average two-point correlation for velocity and magnetic fields {\it vs} $z$ for case C.
  The reference values are taken at  the left $z$ boundary.}
\label{ubcorr}
\end{figure}

\subsection{Analysis of ponderomotive acceleration}\label{sec:apa}

Do we find evidence of effective ponderomotive acceleration in our numerical simulations ?
Figure \ref{pondfvst} shows the maximum ponderomotive acceleration as a function of time for all of the simulated cases.
Note that we have limited the sampling of the ponderomotive acceleration to regions near the $z$ boundaries (the upper chromospheric regions).
In addition, only accelerations away from the $z$ boundaries and toward the center of the simulation
box are included, {\it i.e.,} away from
the chromosphere and toward the corona. This is overwhelmingly the dominant sign of the acceleration.
The maximum ponderomotive acceleration is found to be highly intermittent in time --
its evolution characterized by exhibiting sporadic bursts. Each burst lasts for a characteristic
time equivalent to a few periods of an Alfv\'en wave that would be resonant with loop in the
sense that the wave travel time from one footpoint to the other is equal to an integral multiple of
half a wave period.   In section 4
this period is estimated to be of order 10 s for the loop fundamental.
Note as well that the maximum ponderomotive acceleration often exceeds in magnitude the solar
surface gravitational acceleration (274 m/s$^2$), allowing it to accelerate ions directly into
the corona. However most plasma is probably evaporated from the chromosphere into the corona, with
the ponderomotive acceleration providing only the fractionation, since in realistic conditions
the ponderomotive acceleration is large only over a small range of altitude, where the chromospheric
density gradient is large (see section 4 below). Figure \ref{pondatcorr} supports this point of view, showing
the correlation between the maximum ponderomotive acceleration from case C and the maximum
coronal temperature. Reconnection events that produce the ponderomotive acceleration also increase the
coronal temperature, giving rise to heat conduction downwards and chromospheric evaporation back
upwards.
The length of the loop appears to play a role in the time evolution of the maximum
ponderomotive acceleration
For all of the simulated loop lengths, figure \ref{pondfvst} shows that, after an
interval associated with the loop energization phase, the maximum ponderomotive
acceleration fluctuates around a value of approximately $10^3$ m/s$^2$.  However,
the height of the bursts in activity appears to be correlated inversely with loop length.
\begin{figure}
  \includegraphics[width=1.0\columnwidth, bb = 150 40 635 500]{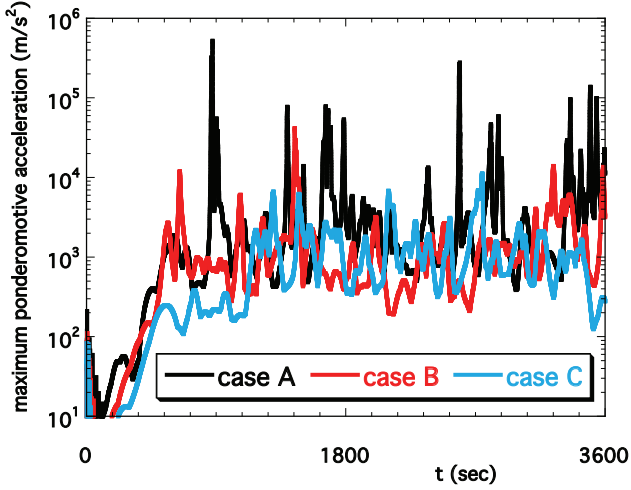}
  \caption{Maximum ponderomotive acceleration {\it vs} time ($t$) for all cases A, B, C.
  This graph shows how the ponderomotive acceleration varies with loop length.
  Note that the maximum ponderomotive acceleration often exceeds the sole surface gravity
  value of 274 m/s$^2$.}
\label{pondfvst}
\end{figure}

\begin{figure}
  \includegraphics[width=1.0\columnwidth, bb = 150 40 635 500]{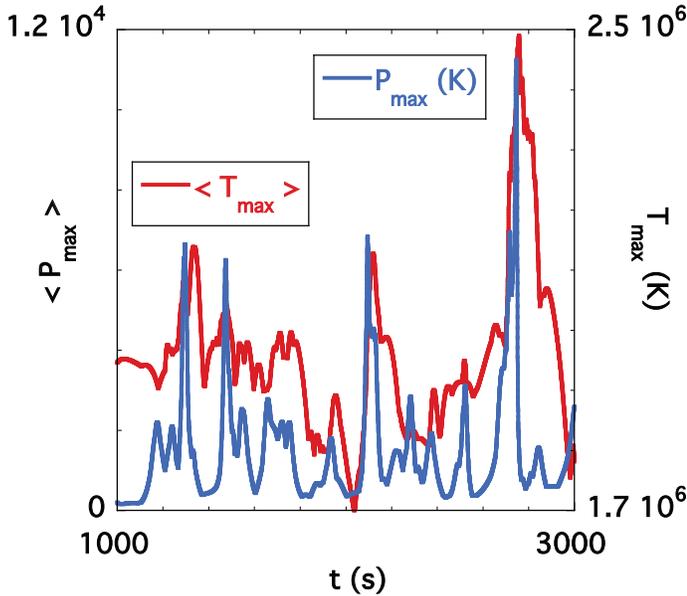}
  \caption{Correlation in time between maximumn coronal temperature and maximum chromospheric
  ponderomotive acceleration for case C.}
\label{pondatcorr}
\end{figure}

\begin{figure}
  \includegraphics[width=1.0\columnwidth, bb = 150 40 635 500]{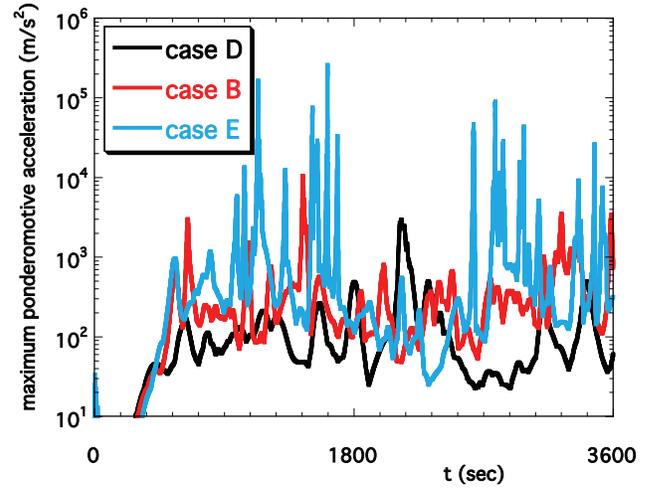}
  \caption{Maximum ponderomotive acceleration {\it vs} time ($t$) for all cases D, B, and E.
    This graph shows how the ponderomotive acceleration varies with magnetic field strength.
  Note that the maximum ponderomotive acceleration often exceeds the sole surface gravity
  value of 274 m/s$^2$.}
\label{pondb}
\end{figure}

A more detailed statistical analysis of the perpendicular dynamical components responsible for the
ideal component of the fluctuating electric field confirms the notion that the
velocity field and not the magnetic field is responsible for the ponderomotive acceleration.
Due to the periodicity in $x$ and $y$ these components all have zero mean values in $z$.
Hence we look at the turbulence intensities, {\it i.e.,} the average root mean square values.
Since there is no mean value, the $u$ turbulence intensity, for example, is given by:
\begin{equation}
<u^2>= {1\over (m_f - m_i)}\sum_{m=m_i}^{m_f}\Bigl[{1\over{n_x}{n_y}}\sum_{j=1}^{n_y}\sum_{i=1}^{n_x} u^2_{ijm}\Bigr].
\end{equation}
The perpendicular velocity turbulence intensities vary significantly from the bases of the loop to the center of the loop (Figure \ref{tiv}),
reflective of the change from the boundary driving motions to the reconnection jets in the system center.
In contrast, the perpendicular magnetic field turbulence intensities are almost constant along the loop (Figure \ref{tib}), with a
value of about one percent of the DC magnetic field in $z$.

The resistive component of the fluctuating electric field is dominated by the fieldwise electric current, which is approximately the
same as $j_z$.
The root mean square $z$ electric current density is shown in Figure \ref{jrms}, from which it can be inferred that
the average resistive component is almost constant along the loop.
Hence the resistive component will not contribute to the electric field gradient in $z$.
A gradient in the fluctuating electric field is required to produce the ponderomotive acceleration.
Thus inspection of the perpendicular turbulence intensities and the resistive component
indicates that the fluctuating velocity field is responsible for the ponderomotive acceleration
since it alone exhibits variation in $z$.

\begin{figure}
  \includegraphics[width=1.\columnwidth, bb = 150 40 635 500]{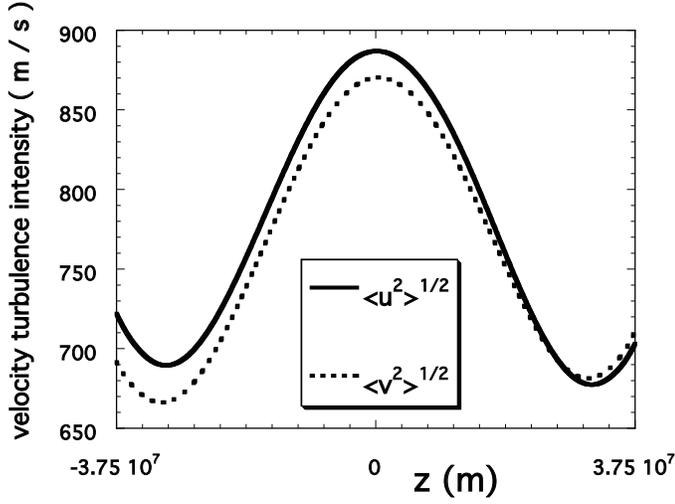}
  \caption{Velocity turbulence intensities  {\it vs} $z$ for case C.
  The velocity turbulence intensities exhibit  a variation of approximately 28$\%$ from
  the boundary to the loop apex. }
\label{tiv}
\end{figure}

\begin{figure}
  \includegraphics[width=1.\columnwidth, bb = 150 40 635 500]{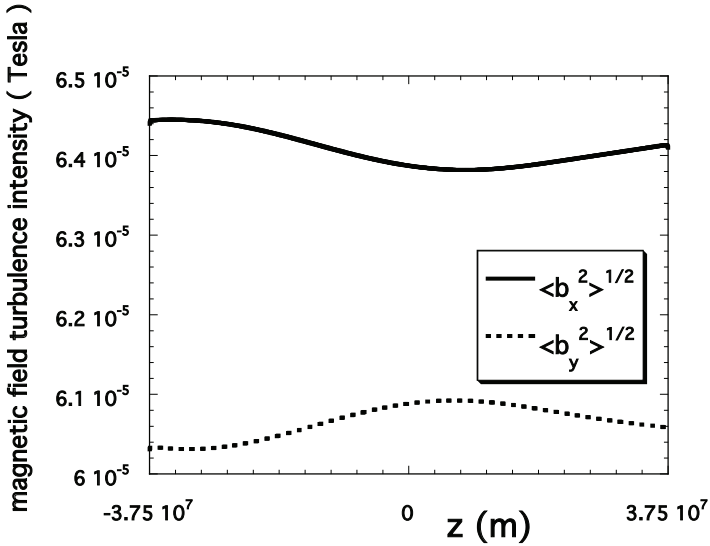}
  \caption{Magnetic field turbulence intensities {\it vs} $z$ for case C.
  The magnetic field turbulence intensities exhibit  a variation of approximately 2$\%$ from
  the boundary to the loop apex.}
\label{tib}
\end{figure}

\begin{figure}
  \includegraphics[width=0.9\columnwidth, bb = 150 40 635 500]{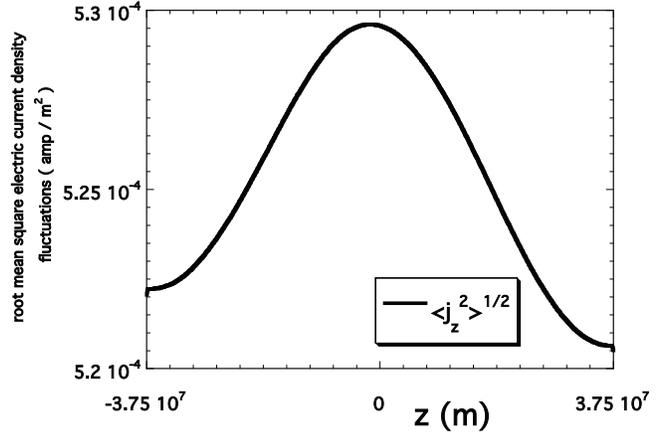}
  \caption{Root mean square electric current density fluctuations {\it vs} $z$ for case C.
  The root mean square electric current fluctuations exhibit  a variation of approximately 2$\%$ from
  the boundary to the loop apex.}
\label{jrms}
\end{figure}

For the shorter loop simulations the perpendicular magnetic turbulence intensities and the
root mean square $z$ electric current density are also relatively unchanged with respect to $z$.
As can be inferred from figure \ref{ubcorr}, the perpendicular velocity intensities exhibit greater
variation in $z$.
For example, figure \ref{tiva} shows the perpendicular velocity turbulence intensities for case A.
The loop apex values for this case are approximately 300 m/s larger than those for case C.
\begin{figure}
  \includegraphics[width=1.\columnwidth, bb = 150 40 635 500]{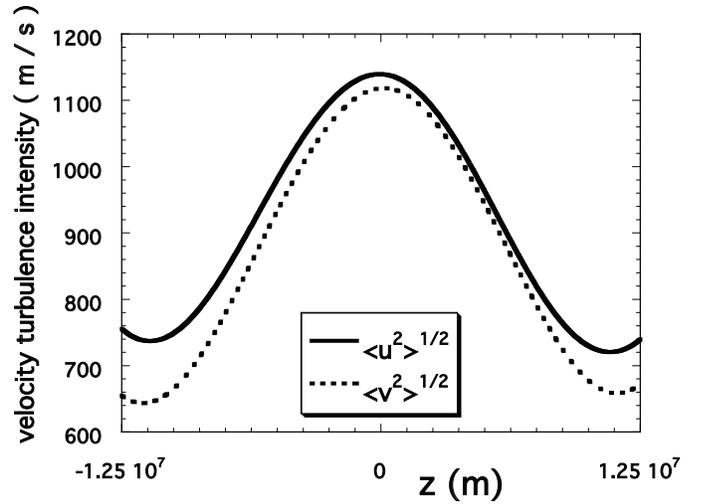}
  \caption{Velocity turbulence intensities  {\it vs} $z$ for case A.
  The velocity turbulence intensities exhibit  a variation of approximately 64$\%$ from
  the boundary to the loop apex.}
\label{tiva}
\end{figure}

\begin{figure}
  \includegraphics[width=1.0\columnwidth,]{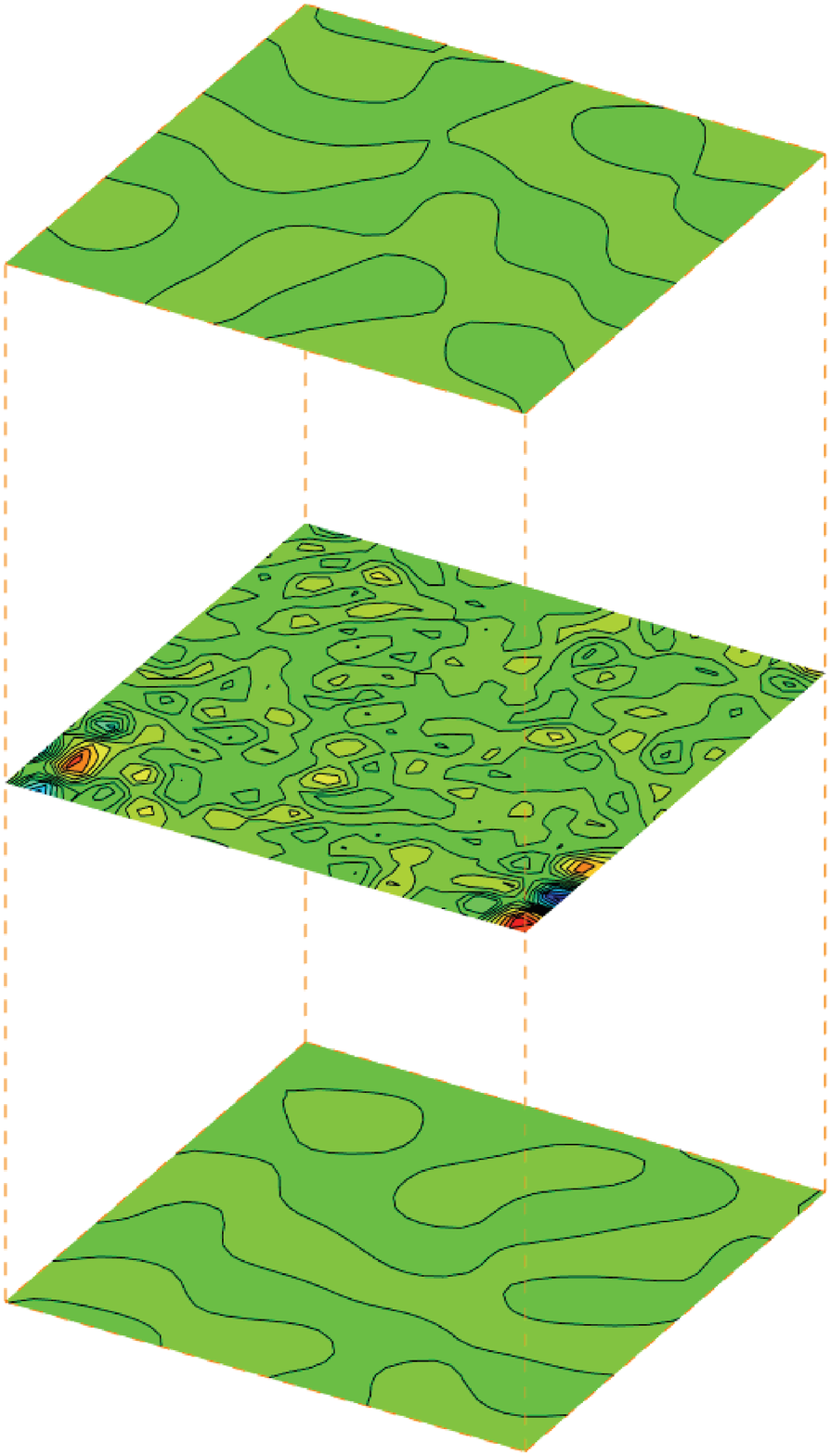}
  \caption{The $z$ vorticity at the walls and the channel center.  The aspect ratio of the computational box
  has been altered to enhance the visualization.  The structure at the boundaries is due to the forcing flow.
   The structure in the interior is related to the formation of magnetic reconnection jets and magnetohydrodynamic
   turbulence.}
\label{vortcut}
\end{figure}
It is the perpendicular velocity field components that interest us the most because these have the most effect on the fluctuating electric field since they are crossed into the DC magnetic field. {\it i.e.} the large magnetic field in the $z$-direction (equation \ref{eeq}).
The average perpendicular velocity field varies in a way appropriate for ponderomotive accelerations away from the $z$-boundary, i.e. the reconnection jet velocity is larger than the footpoint driving velocity. This variation in the flow field is caused by the mismatch between the $z$-boundary physics and the $z$-midplane physics. At the wall we have specified a large scale, incompressible flow field (see subsection 2.2). In the vicinity of the $z$-midplane the flow is dominated by jets caused by magnetic reconnection. The jets have a much higher velocity and much smaller spatial scale than the
$z$-boundary flows. This is apparent when the vorticity at the $z$-boundaries is compared with the vorticity at the $z$-midplane (Figure \ref{vortcut}). There is a transitional region where these two flows are blended. It is in this region that the ponderomotive acceleration occurs.

Figure \ref{pondfvst} showed that the loop ponderomotive acceleration is temporally intermittent.
It's likely then that the turbulent fields are spatially intermittent as well.
A look at some higher-order statistical quantities will help to determine whether this is so.
The time-averaged skewness factor profile (figure \ref{skew}) is given by:
\begin{equation}
S(f)={<f^3>\over<f^2>^{3/2}}
\end{equation}
and the time-averaged kurtosis factor profile (figure \ref{kurt}) is given by:
\begin{equation}
K(f)={<f^4>\over<f^2>^2}.
\end{equation}
Nonintermittent turbulence is characterized by a skewness of zero and a kurtosis of three.
This set of values produces a skewed mesokurtic distribution of fluctuations -- commonly called a Gaussian or normal distribution.
Departures from these values indicate various types of intermittent turbulence.
From Figure \ref{skew} and Figure \ref{kurt} we conclude that, approximately speaking, the perpendicular magnetic field has an unskewed mesokurtic distribution at both the walls and the center of the channel.
The perpendicular velocity field approximately has an unskewed mesokurtic distribution
at the walls and an unskewed leptokurtic distribution in the channel.
A leptokurtic distribution implies greater excursions about the mean than for the mesokurtic
distribution, {\it i.e.,} the quantity being analyzed will have a distribution with values which cluster about the mean, leading to a higher peak, and also
have occasional large excursions away from the mean, leading to a distribution with thicker tails.
This intermittent distribution for the perpendicular velocities near the loop apex arises as a consequence of the formation of reconnection jets in the center of the channel.
The variations are negligible in the system interior except where magnetic reconnection occurs.
To use the sensor picture described earlier (see section \ref{sec:sensor}), at one of the base sensors the imposed $x$-velocity, for example, moves smoothly between positive and negative values as measured in time as determined by boundary convection pattern.
In the loop, however, the $x$-velocity generally remains close to zero.  However, when a reconnection event occurs, the $x$-velocity exhibits sporadic positive or negative bursts in time, evidence of concentrated, high speed structures.
The same remarks apply to the $y$ velocity.
This variation in physical processes leads to the $z$ variation in velocity kurtosis.

\begin{figure}
  \includegraphics[width=1.\columnwidth, bb = 150 40 635 500]{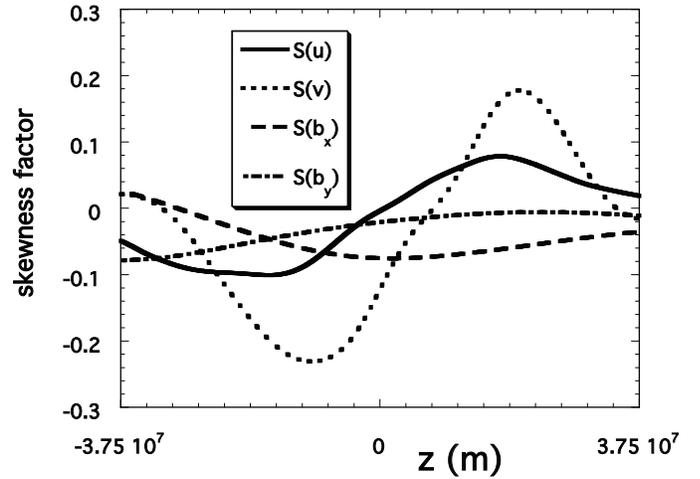}
  \caption{Velocity field and magnetic skewness factors {\it vs} $z$ for case C.}
\label{skew}
\end{figure}

\begin{figure}
  \includegraphics[width=1.\columnwidth, bb = 150 40 635 500]{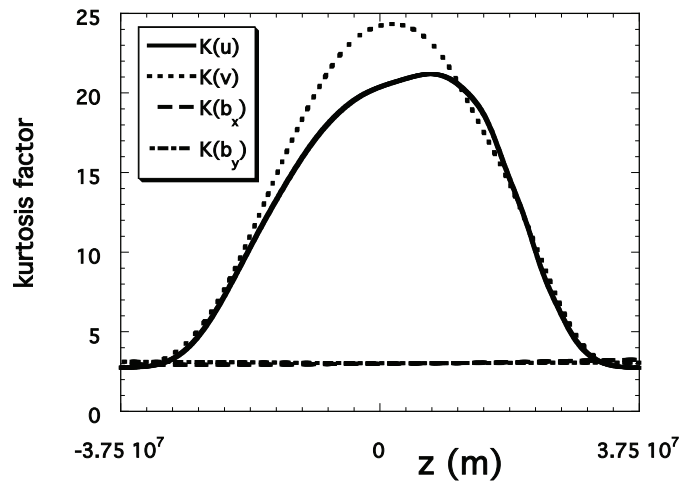}
  \caption{Velocity field and magnetic kurtosis factors {\it vs} $z$ for case C.}
\label{kurt}
\end{figure}

Figure \ref{pondfvst} indicates that the shorter loop cases have a higher degree of temporal intermittency.
Does this behavior have a counterpart in the skewness and kurtosis for the shorter loop cases ?
Figure \ref{skewb} shows the skewness factors for case B.
The perpendicular magnetic skewness  and velocity skewness factors again remain close to zero.
This behavior finds a counterpart in the kurtosis factors, which are shown for case B in Figure \ref{kurtbb}.
For the perpendicular magnetic field the kurtosis factor again remains close to a value of 3.
The perpendicular velocity field kurtosis factor near the loop apex for case B has increased
relative to the value that it has for case C.

\begin{figure}
  \includegraphics[width=1.\columnwidth, bb = 150 40 635 500]{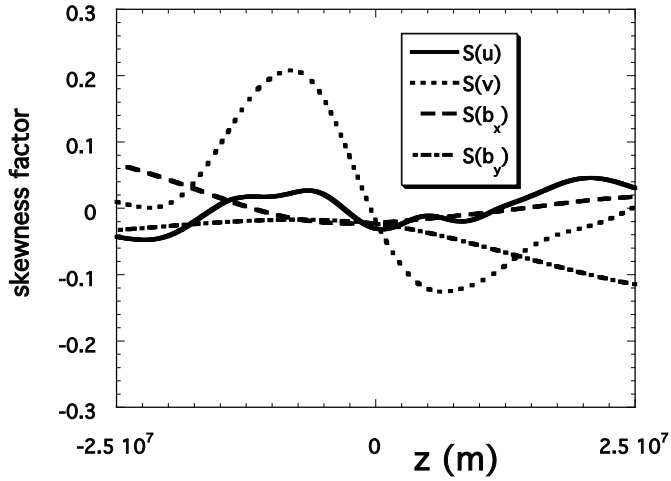}
  \caption{Velocity field and magnetic skewness factors {\it vs} $z$ for case B.}
\label{skewb}
\end{figure}

\begin{figure}
  \includegraphics[width=1.\columnwidth, bb = 150 40 635 500]{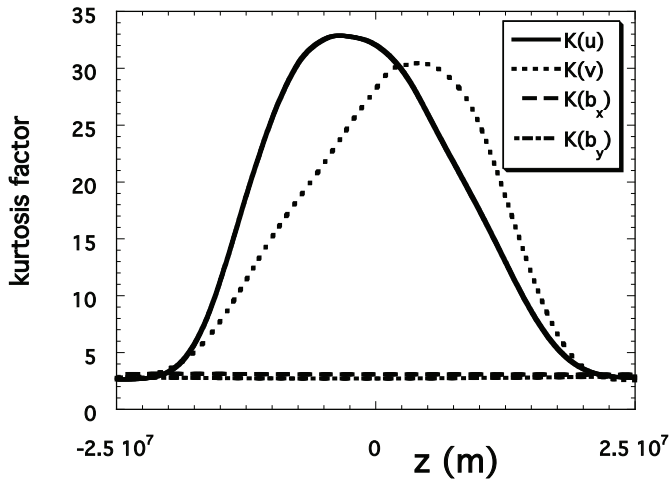}
  \caption{Velocity field and magnetic kurtosis factors {\it vs} $z$ for case B.}
\label{kurtbb}
\end{figure}

Figure \ref{pondb} indicates that the higher magnetic field cases have a higher degree of temporal intermittency.
This behavior again finds a counterpart in the kurtosis factors, which are shown for case E in Figure \ref{kurtb}.
For the perpendicular magnetic field the kurtosis factor again remains close to a value of 3.
The perpendicular velocity field kurtosis factor near the loop apex for case E has increased
relative to the value that it has for case B.

\begin{figure}
  \includegraphics[width=1.\columnwidth, bb = 150 40 635 500]{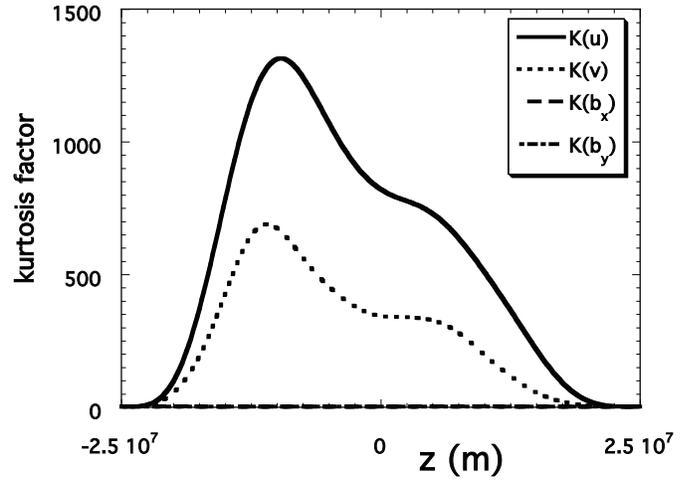}
  \caption{Velocity field and magnetic kurtosis factors {\it vs} $z$ for case E.}
\label{kurtb}
\end{figure}

Figure \ref{pondpdf} shows regions of probability density function at $t=1800$ seconds for case C
where the ponderomotive acceleration exceeds the solar surface gravity (274
m/s$^2$). The low values for the probability show that only a very small part
of the loop is likely to be involved in the FIP effect, {\it i.e.,} the sections of the loop that are close to
the chromospheric boundaries.
\begin{figure}
  \includegraphics[width=1.0\columnwidth, bb = 150 40 635 500]{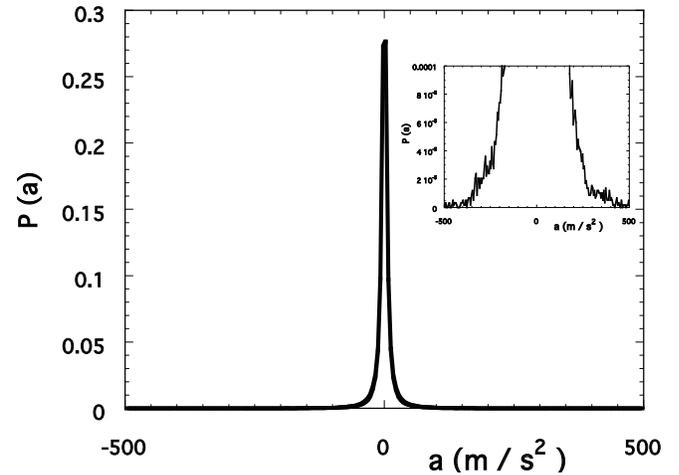}
  \bigskip
  \caption{Probability density function of the ponderomotive acceleration for case C.
                Inset highlights regions where ponderomotive acelleration exceeds solar surface gravity.}
\label{pondpdf}
\bigskip\bigskip\bigskip
\end{figure}


\section{Discussion}
The ponderomotive acceleration is usually discussed in connection with Alfv\'en or fast mode waves.
In homogeneous media, the ponderomotive acceleration associated with slow mode waves
depends on the negative gradient of wave pressure. Being communicated through particle collisions
rather than interactions with electric or magnetic fields, it does not cleanly separate ions from
neutrals in the way that the ponderomotive acceleration due to Alfv\'en or fast modes does. It is
also smaller by one to two orders of magnitude than
the ponderomotive acceleration considered here, comparable with the acceleration due to the thermal
force but with a different mass dependence. Exceptions may occur with slow mode waves in thin flux tubes,
where a transverse velocity perturbation, and hence corresponding field perturbations, may exist
\citep[e.g. the ``sausage'' mode in][]{mikhalyaev05} but such considerations are beyond the scope
of the current paper. In addition, the longitudinal pressure associated with slow mode waves also
acts to inhibit
fractionation \citep[see equations 3 and 23 in][]{laming12}. Throughout this work we also make the
approximation that when the longitudinal slow mode wave amplitudes added in quadrature are greater
than the local sound speed, turbulence associated with the shock that develops shuts off all further
fractionation.

In this work it has not been possible to identify the various MHD waves directly,
 due to the complexity of
the simulations, and we have focused on the
coronal velocity field produced by reconnection jets. Here we briefly revisit the wave picture.
Waves can be
either generated directly at the
reconnection site, or excited as the reconnection jet interacts with surrounding plasma.
\citet{sturrock99} gives a pedagogic
review of the mechanisms by which various wave modes may be excited by
reconnection. The reconnected
field line is generally distorted, and this can either propagate away
from the reconnection site as an Alfv\'en wave, or emit
magnetoacoustic waves traveling perpendicularly to the magnetic field
direction  (the only MHD wave that can propagate in this direction).
\citet{kigure10} consider the generation of Alfv\'en waves
by magnetic reconnection in a more quantitative fashion,
and find that a significant fraction of the magnetic energy released
(several tens of \%, depending on geometry and plasma $\beta$) can be
carried off by Alfv\'en or magnetoacoustic (fast or slow mode) waves,
with Alfv\'en waves dominating for $\beta <1$.

According to \citet{sturrock99}, higher frequency waves may be associated
with ``plasmoids'' or magnetic islands in the current sheet. This so-called
plasmoid instability has attracted interest because of its role in mediating
fast reconnection \citep[e.g.][]{ni15}. \citet{Loureiro07} have shown that
the criteria for onset of the plasmoid instability is that the plasma
Lundquist number $S\equiv(\mu _0L V_A/\eta)$, has to be greater than $\approx
10^4$.  Further work by \citet{Bhattacharjee09}, \citet{Lapenta08}, and
\citet{Daughton09} demonstrated plasmoid generation and $S$-independent or
weakly dependent reconnection rates for $S > 10^4$. This condition is likely
to be easily met in coronal nanoflares as in this paper (see Table 1), where
$\eta$ is low and
$V_A$ is high, and observational signatures of drifting and pulsating
structures in solar flare associated current sheets have been interpreted by
\citet{Kliem00} and \citet{Karlicky04} as evidence of embedded secondary
plasmoids. \citet{oishi15} find no dependence of reconnection rate on $S$ in
a range $3.2\times 10^3 \le S \le 3.2\times 10^5$, and argue that other
turbulence besides the plasmoid instability must dominate the reconnection
rate.

Waves may also arise from the interaction of the reconnection jet with
ambient plasma, or it may evolve to become unstable itself. \citet{hoshino15}
investigate the streaming tearing and sausage modes, and the streaming kink
mode. This last mode has the highest growth rate, approximately independent
of $S$. \citet{liu11} discuss the role of temperature anisotropies and wave
generation by the firehose instability in the outflow, with increasing
firehose instability with increasing obliquity, i.e. at higher guide fields.

The ponderomotive acceleration was found to be strongly intermittent, as
would be expected if due to waves released by reconnection. In our
simulations, the Alfv\'en speed $V_A\simeq 4\times 10^6\left(B/{\rm
0.01T}\right)/\sqrt{n/\left(3\times 10^{15} {\rm m}^{-3}\right)}$ m s$^{-1}$ is comparable
to the electron thermal speed, meaning that following energy release in the
corona, Alfv\'en waves are likely to arrive at the chromosphere and
fractionate the plasma before significant heat conducted down can cause an
evaporative upflow. The magnitude of the acceleration can be sufficient to
overcome solar surface gravity. In fact Fig. 4 shows that it is much higher
than this for much of the period of the simulation. Previous work
\citep{laming15} has shown that a ponderomotive acceleration in the steep
density gradient of the solar chromosphere of about $10^4$ m s$^{-1}$
reproduces the observed solar FIP fractionation. This is illustrated further
in Fig. \ref{chromo}, which shows the wave propagation and FIP fractionation in the
chromosphere of a 50,000 km loop with $B=0.01$ T, designed to match case B.
The Alfv\'en wave transport equations are solved for parallel propagating
undamped waves, with angular frequency $\omega =0.590$ rad s$^{-1}$ which
places them on resonance with the coronal loop ($\omega = V_A/2L_z$). The top
left panel shows the Els\"asser variables, $\delta v_{\perp}$ and $\delta
B_{\perp}/\sqrt{4\pi nm_p}$. Real parts are in black, imaginary parts in
gray. The chromosphere has a steep density gradient at an altitude of about
2.285 $\times 10^7 $m, which is reflected in steep gradients in the wave amplitudes.

The steep chromospheric density gradient also gives rise to a strong
``spike'' in the ponderomotive acceleration, with maximum value about $10^4$
m s$^{-1}$, shown in the bottom left panel as a solid line. The dotted line
here gives the amplitude of slow mode waves generated parametrically by the
incident coronal Alfv\'en wave. The bottom right panel shows the FIP
fractionations (black lines) for the element abundance ratios Si/O, C/O,
Fe/O, Mg/O, He/O, and S/O, calculated from equation 22 in \citet{laming15}
using the computed ponderomotive acceleration profile. Fe/O increases by a
factor of about 3.7 and Mg/O by about 3.2, comparable to solar observations.
Other elements less highly ionized in the chromosphere, e.g. S and C, are
fractionated by much less, and He and Ne, the elements with the highest first
ionization potentials are depleted relative to O. The chromospheric
ionization fractions for each element are shown in gray, to be read on the
right hand side axis, and are displayed on an expanded plot in the top right
panel.

The ponderomotive acceleration in our simulations varies much more
dramatically than the observed solar FIP fractionation, and develops in the shallower
background density gradient, potentially developing a stronger fractionation. In fact one of the
curious things about the FIP effect is that, globally at least, it is
surprisingly constant at a factor of 3-4 enhancement, not often more or less. This
leads us to speculate about mechanisms that are either absent from the
simulations, or inadequately represented, that might regulate or reduce
the degree of fractionation.

\citet{laming15} suggests that the ponderomotive acceleration may begin to
change the chromospheric density structure, reducing the density gradients
and diminishing the ultimate fractionation. Although the Newton cooling
function used near the coronal base may limit this aspect of the simulations
in this region, more recent estimates following \citet{laming15} suggest that
this is not the sole agent of saturation of the FIP effect, and that other
processes such at wave damping must be at work.

A large amplitude coronal Alfv\'en wave restricted to a small part of the
loop cross section will inevitably give rise to regions of strong velocity
shear where drift waves may be excited \citep[e.g.][]{vranjes10a,vranjes10b}.
Such waves require non-zero ion gyroradii and are thus not treated in the
HYPERION fluid simulations. Other modes of wave damping like turbulent
cascade are included, but possibly not fully treated due to the limitations
of spatial resolution necessary in a 3D simulation. \citet{demoortel14} and
\citet{liu14} see strong damping of presumably Alfv\'en waves towards the
apex of coronal loops, sufficiently strong that in each loop leg waves are
only seen to propagate in one direction. Such damping suggests that high
amplitude waves must exist on the loop, and that such waves are perhaps less likely to be
torsional waves which do not cascade very effectively \citep{vasheghani12}.

Further treatment of these and other effects is beyond the scope of this
work. We reiterate our main point. A coronal loop subject to heating by
the continuous formation and dissipation of field-aligned current sheets
naturally produces a ponderomotive acceleration at the loop footpoints
adequate to give rise to the observed solar FIP fractionation. This supports
the model proposed by \citet[][and references therein]{laming15} whereby the
FIP effect arises from the action of the ponderomotive force on chromospheric
ions which are selectively accelerated up into the corona, and that the waves
responsible for the ponderomotive force should have a coronal origin.

\section{Conclusions}
In this paper we have examined the development of ponderomotive acceleration in a
coronal loop threaded by a
strong axial magnetic field whose footpoints are convected by random motions.
The loop develops highly dynamical behavior that leads to the formation of magnetic
and flow structures associated
with magnetic reconnection.
The energy release is highly intermittent in space and time, and thus the magnetic
and velocity structures exhibit considerable spatial and temporal variation as well.
It is the interaction of the forcing flow and the magnetic
reconnection flow that produces ponderomotive acceleration.
This occurs near the loop footpoints at both $z$ boundaries.
A statistical analysis of components of the ponderomotive acceleration affirms that variations
in the velocity field are its most important cause.

The key point of this paper is that ponderomotive acceleration is a consequence of
coronal heating.
It therefore occurs ubiquitously in coronal loops and no special process needs to be
invoked for its presence.
Rather, as long as current sheets and magnetic reconnection jets form in response to
footpoint motions ponderomotive acceleration will occur.

A preliminary survey of the variation of ponderomotive acceleration with loop length and
magnetic field strength was conducted.  It was found that ponderomotive acceleration
occurs in all of the loops tested.
In general, over the limited range being tested, the magnitude of ponderomotive acceleration
was found to decrease with loop length and increase with magnetic field strength.
Changes in intermittency were also observed.  Higher peak acceleration values were found
for loops with shorter lengths of larger magnetic fields.  It would be desirable to extend the
range of values in this survey.

\acknowledgments
RBD thanks G. Einaudi and A. F. Rappazzo for helpful discussions.
This research was supported by NRL 6.2 funds and by the NASA SR\&T program.
Computer simulations were performed on the LCP\&FD Intel Core17 cluster.

\begin{figure}[t]
  \includegraphics[width=7truein]{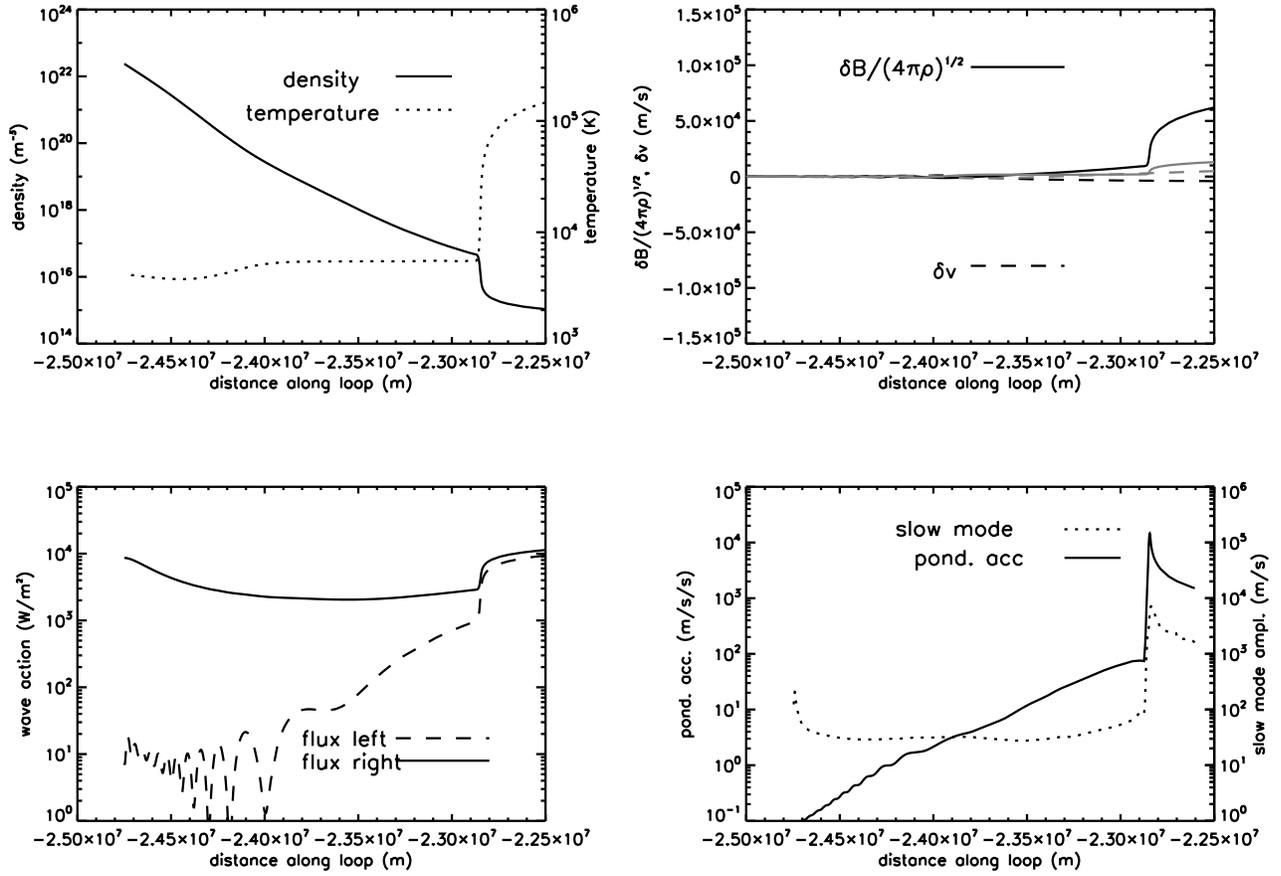}
  \bigskip
  \caption{Chromospheric portion of solution of Alfv\'en wave transport equations for coronal
  loop in Fig. 1. The top left panel shows the density and temperature profile in the chromosphere.
  Top right shows the Els\"asser variables, bottom left the wave energy fluxes and bottom right the
  ponderomotive acceleration and associated slow mode waves resulting from parametric generation. The
  steep density gradient at about $-2.285\times 10^7$ m shows up in the behavior of all the wave
  variables in the other three panels.}
\label{chromo}
\bigskip\bigskip\bigskip
\end{figure}

\begin{figure}[t]
  \includegraphics[width=7truein]{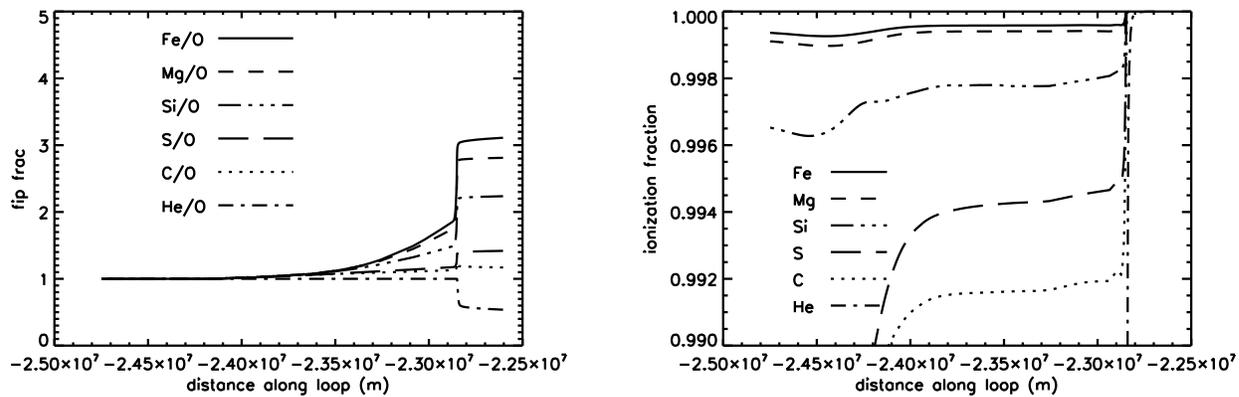}
  \bigskip
  \caption{The FIP fractionations relative to O resulting from the ponderomotive acceleration (left panel).
  For reference, the ionization fractions of the various elements in the model chromosphere are
  shown at right.}
\label{chromo_2}
\bigskip\bigskip\bigskip
\end{figure}

\appendix

\section{Governing equations}
The governing equations, written here in dimensionless form, are:
\begin{eqnarray}
{{\partial n}\over {\partial t}} &=& -\nabla\cdot (n {\bf v}),  \label{eq:eqn}      \\[.4em]
{{\partial n {\bf v}}\over{\partial t}} &=& -\nabla\cdot({n \bf v v})
   -{\beta}\nabla p + {\bf J}\times{\bf B}+
{1\over S_v}\nabla\cdot{\bf \zeta} \nonumber
+ \frac{1}{Fr^2}\   n \Gamma(z)\, {\bf\hat e}_z \label{eq:eqnv} \\[.4em]
{{\partial T}\over{\partial t}} &=& -{\bf v}\cdot\nabla T
 - (\gamma - 1) (\nabla\cdot {\bf v}) T
\nonumber
+\frac{1}{n} \Bigg\{ \frac{1}{ Pr\, S_v }
\bigg[{\bf B}\cdot\nabla
\bigg( \kappa_{\parallel}\ T^{5/2}\ {{\bf B}\cdot\nabla T\over B^2}\bigg) \bigg]
\nonumber \\
&& + {(\gamma -1)\over\beta}  \bigg[
{ 1\over S_v} \zeta_{ij} {\partial v_i\over\partial x_j}
+{1\over S} (\nabla\times{\bf B})^2
 -{1\over P_{rad} S_v} n^2\Lambda (T)
+ {\beta\over(\gamma - 1)} n C_N \bigg] \Bigg\}, \label{eq:eqT}\\[.4em]
{{\partial {\bf B}}\over{\partial t}} &=& \nabla\times{\bf v}\times{\bf B}
  - \frac{1}{S}\nabla\times \nabla\times {\bf B}, \label{eq:b}\\[.4em]
&&\nabla\cdot{\bf B} = 0, \label{eq:divb}\\[.4em]
\nonumber
\end{eqnarray}
\noindent
and
\begin{equation}
p = nT. \label{eq:eqp}
 \end{equation}

The non-dimensional variables are defined in the following way:
$n ({\bf x}, t)$ is the number density,
${\bf v}({\bf x}, t) = (u, v, w)$ is the flow velocity,
$p({\bf x}, t)$ is the thermal pressure,
${\bf B}({\bf x}, t) = (B_x, B_y, B_z) $ is the magnetic induction field,
${\bf J} = \nabla\times{\bf B}$ is the electric current density,
$T({\bf x}, t)$ is the plasma temperature,
$\zeta_{ij}= \mu (\partial_j v_i + \partial_i v_j) -
\lambda \nabla\cdot {\bf v} \delta_{ij}$ is the viscous stress tensor,
$e_{ij}= (\partial_j v_i + \partial_i v_j)$ is the strain tensor,
and $\gamma$ is the adiabatic ratio.

To render the equations dimensionless we set characteristic values at the
walls of the computational box: a number density $n_*$,
vertical Alfv{\'e}n speed at the boundaries $V_{A*}$,
the orthogonal box width $L_*$,  and the temperature $T_*$.
Time ($t$) is measured in units of the Alfv\'en time
($\tau_A=L_* /V_{A*}$).

The magnetic resistivity $\eta$, and shear viscosity $\mu$
are assumed to be constant and uniform, and Stokes relationship is assumed
so the bulk viscosity $\lambda = (2/3) \mu$.
The parallel thermal conductivity is given by $\kappa_\parallel$,
while the perpendicular thermal conduction ($\kappa_{\perp}$) is set to zero.
We use the radiation function based on the CHIANTI atomic database
\citep{2012ApJ...744...99L}, normalized by its value at the base temperature
$T_* = 10000\, K$.
The loop gravity profile $[\Gamma (z)]$ is determined by an elliptical model
\citep{2015ApJ...submitted}.
The term $C_N$ denotes a Newton cooling function which is enforced
near the loop base \citep{2001A&A...365..562D}.
We use $C_N = 10~[T_i (z) - T(z)] e^{-4(z+0.5L_z)}$ at the lower boundary and
$C_N = 10~[T_i (z) - T(z)] e^{-4(0.5L_z-z)}$ at the upper boundary,
where $T_i (z)$ is the initial temperature profile.

Thus the important dimensionless numbers are:
$S_v = n_* m_p  V_{A*} L_* / \mu \equiv$ viscous Lundquist number
($m_p = 1.673\times 10^{-27}$ kg is the proton mass),
$S = \mu_0 V_{A*} L_* / \eta \equiv$ Lundquist number
($\mu_0 = 1.256\times10^{-6}$ Henrys / meter is the magnetic permeability),
$\beta = \mu_0 p_* / B_*^2 \equiv$ pressure ratio at the wall,
$Pr = C_v \mu / \kappa_{\parallel} T_*^{5/2} \equiv$ Prandtl number, and
$P_{rad} $, the radiative Prandtl number
${\mu/ \tau_A^{2} n_*^2 \Lambda (T_*)} $. $C_v$ is
the specific heat  at constant volume.
The magnetohydrodynamic Froude number ($Fr$) is equal to $V_A/(g L_*)^{1/2}$,
where $g=274$~m~s$^{-2}$ is the solar surface gravity.

\section{Numerical method}

With the previous definitions, equations ~\ref{eq:b}  and \ref{eq:divb} can be replaced by the
magnetic vector potential equation:
\begin{equation} \label{eq:eqbp}
 {\partial {\bf A}\over\partial t}={\bf v}\times ( B_0\, \mathbf{\hat e}_z + \nabla\times{\bf A} )
  - {1\over S}~ \nabla\times\nabla\times {\bf A}
\end{equation}
We solve numerically the equations~\ref{eq:eqn}-\ref{eq:eqT} and \ref{eq:eqbp}
together with equation~\ref{eq:eqp}.  Space is discretized in $x$ and
$y$ with a Fourier collocation scheme \citep{1989PhFlB...1.2153D} with isotropic
truncation dealiasing.
A second-order central difference technique is used for the
discretization in $z$ \citep{1986JFM...169...71D}.
Variables are advanced in time by a five--step--fourth-order
low-storage Runge-Kutta scheme\citep{carpenter1994fourth}.
Thermal conduction is advanced with second-order Super TimeStepping
\citep{2012MNRAS.422.2102M}.
HYPERION employs a hybrid parallelization using a combination of OpenMP and MPI.  For the results presented in this paper, HYPERION was run on a cluster of two-socket Intel Xeon X5650 (Westmere) nodes, with one MPI rank per socket and six OpenMP threads per MPI rank.

\section{Simulation rescaling}
Lundquist and Prandtl numbers are rescaled to permit resolved calculations, as described in our previous paper \citep{2015ApJ...submitted}.
The simulation parameter values are given in Table 1.






\begin{table*}
\begin{center}

\caption{\label{tab:table 1}  Dimensionless numbers used in simulations. }
\bigskip

\begin{tabular*}{\textwidth}{c @{\extracolsep{\fill}} ccccccc}
\hline \hline\noalign{\vspace{.5em}}
 Case & Length (km) &  $\beta$ & $S_v$ & $S$ & $Fr$  & $Pr$  & $P_{rad}$ \\[.6em]
\hline\noalign{\vspace{.5em}}
A&25000&  $1.735 \times 10^{-4}$&$1.379\times 10^5$& $1.780\times 10^5$ & $2.083\times 10^1$ & $3.835\times 10^{2}$ & $1.111\times10^{-2}$ \\[.3em]
B&50000&  $1.735 \times 10^{-4}$&$1.379\times 10^5$& $1.780\times 10^5$ & $2.083\times 10^1$ & $3.835\times 10^{2}$ & $1.111\times10^{-2}$ \\[.3em]
C&75000&  $1.735 \times 10^{-4}$&$1.379\times 10^5$& $1.780\times 10^5$ & $2.083\times 10^1$ & $3.835\times 10^{2}$ & $1.111\times10^{-2}$ \\[.3em]
D&50000&  $6.942 \times 10^{-4}$&$6.896\times 10^4$& $8.898\times 10^5$ & $1.041\times 10^1$ & $3.835\times 10^{2}$ & $2.778\times10^{-3}$ \\[.3em]
E&50000&  $4.333 \times 10^{-5}$&$2.758\times 10^5$& $3.559\times 10^5$ & $4.166\times 10^1$ & $3.835\times 10^{2}$ & $4.444\times10^{-2}$ \\[.3em]
\end{tabular*}

\end{center}
\end{table*}

\end{document}